\documentclass[10pt,twocolumn,letterpaper]{article}

\usepackage[pagenumbers]{cvpr} 
\usepackage{multirow}
\usepackage{multicol}
\usepackage{color}
\usepackage{xcolor}
\usepackage{mathrsfs}
\usepackage{makecell}
\usepackage{amsmath}
\usepackage{indentfirst}
\usepackage{placeins}

\newcommand{\bx}{\mathbf{x}}

\definecolor{cvprblue}{rgb}{0.21,0.49,0.74}
\usepackage[pagebackref,breaklinks,colorlinks,citecolor=cvprblue]{hyperref}

\title{DifAugGAN: A Practical Diffusion-style Data Augmentation for GAN-based Single Image Super-resolution}

\author{
Axi Niu$^1$, Kang Zhang$^2$, Joshua Tian Jin Tee$^2$, Trung X. Pham$^2$, Jinqiu Sun$^1$,\\ Chang D. Yoo$^2$, In So Kweon$^2$, Yanning Zhang$^1$ \\
$^1$ Northwestern Polytechnical University\\
$^2$ Korea Advanced Institute of Science Technology\\
{\tt\small $^{1}$\{nax@mail., sunjinqiu@,  ynzhang@\}nwpu.edu.cn \quad }\\
{\tt\small $^{2}$\{zhangkang, joshuateetj, trungpx, cd\_yoo, iskweon77\}@kaist.ac.kr}
}

\begin{document}
\maketitle
\begin{abstract}
It is well known the adversarial optimization of GAN-based image super-resolution (SR) methods makes the preceding SR model generate unpleasant and undesirable artifacts, leading to large distortion. We attribute the cause of such distortions to the poor calibration of discriminator, which hampers its ability to provide meaningful feedback to the generator for learning high-quality images.
To address this problem, we propose a simple but non-travel diffusion-style data augmentation scheme for current GAN-based SR methods, known as DifAugGAN. It involves adapting the diffusion process in generative diffusion models for improving calibration of the discriminator during training motivated by the successes of data augmentation schemes in the field to achieve good calibration.
Our DifAugGAN can be a Plug-and-Play strategy for current GAN-based SISR methods to improve the calibration of the discriminator and thus improve SR performance.
Extensive experimental evaluations demonstrate the superiority of DifAugGAN over state-of-the-art GAN-based SISR methods across both synthetic and real-world datasets, showcasing notable advancements in both qualitative and quantitative results.

\end{abstract}    
\section{Introduction}

Single image super-resolution (SISR) stands as a pivotal pursuit within low-level computer vision, aiming to reconstruct high-resolution (HR) images from their low-resolution (LR) counterparts~\cite{zhang2018image,niu2022ms2net}. Similar to various inverse problems, SISR presents challenges due to the potential existence of multiple HR images that align with a single LR input image with an unknown degradation~\cite{saharia2022image}.

Since the pioneering work of SRCNN, which demonstrated notable performance in super-resolution by introducing a straightforward CNN architecture, CNN-based methods have gradually emerged as the predominant solution for image super-resolution~\cite{lim2017enhanced,zhang2018image,cai2019toward,niu2023gran}. These approaches implicitly tackle the SISR problem by learning the intricate, nonlinear LR-to-HR mapping from numerous LR-HR image pairs. While these CNN-based SISR methods optimize the training process using basic image reconstruction losses (e.g., $L_{1}$ or $L_{2}$), yielding impressive PSNR results, these losses have shown a tendency to drive reconstructed results toward an average of potential SR predictions, often leading to over-smoothing in the images~\cite{wu2021contrastive,wang2022low,wu2023practical,xue2023burst,niu2023learning}.

\begin{figure}[t]
  \centering
  \includegraphics[width=1 \linewidth]{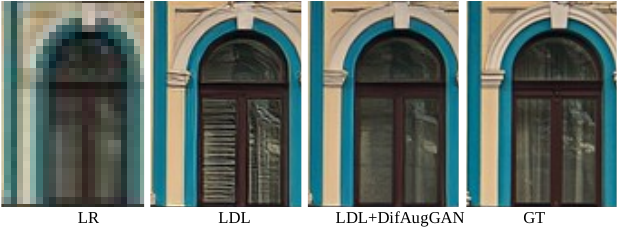}
  \caption{From left to right: LR input, SR  obtained using the artifacts removing oriented method LDL~\cite{liang2022details}, SR result employing our LDL+DifAugGAN, and the ground truth.  Our DifAugGAN notably enhances the performance of LDL, generating more realistic outcomes while significantly reducing distortions.
  }
  \label{fig:artifacts}
\end{figure}

To achieve SR results with better perceptual quality, GAN-based SR methods emerge to meet a historic destiny~\cite{ledig2017photo,wang2018esrgan,zhang2020deep,zhang2021designing,tian2022generative}. GAN-based SISR methods~\cite{wang2018esrgan,soh2019natural,tian2022generative} often involve a generator and a discriminator in an adversarial way to push the generator to generate realistic images. The generator can generate an SR result for the input LR, and the discriminator aims to distinguish if the generated SR is true. The training process is optimized by combining the reconstruction loss and adversarial loss, which in fact helps the GAN-based SR methods recover sharp images with rich details. However, this true-or-false adversarial optimization is vulnerable to generating SR images with a high degree of distortions, as shown in Fig.~\ref{fig:artifacts}. 

\begin{figure}[t]
  \centering
  \includegraphics[width=0.9 \linewidth]{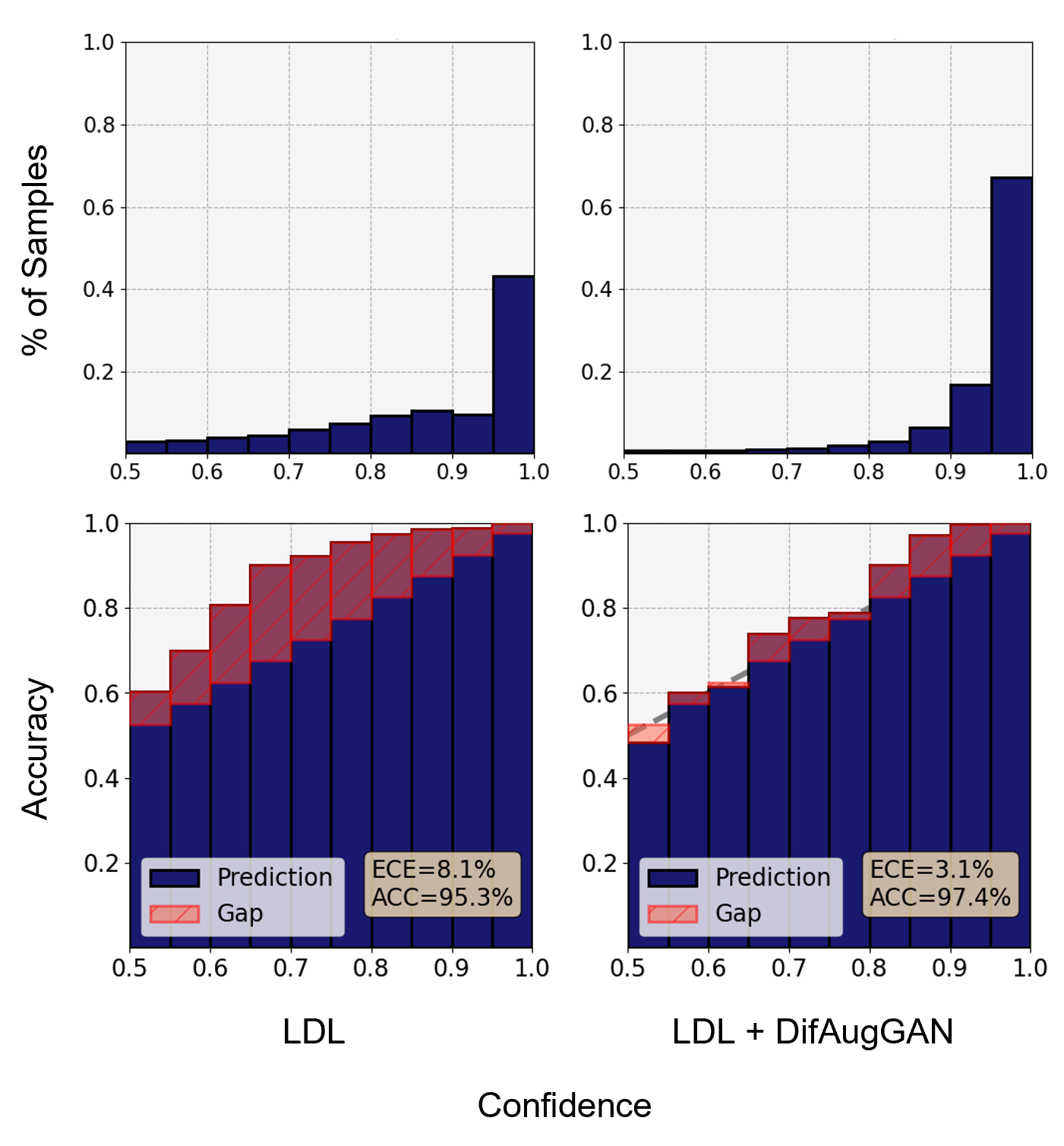}
  \caption{Confidence histograms (top) and reliability diagrams
(bottom) comparing LDL (left) with LDL+DifAugGAN (right) on URBAN100. The top row illustrates the ratio of samples predicted with corresponding confidence, while the bottom row visualizes the alignment of the model prediction with the gap between an ideal calibration model. Evidently, the introduction of our DifAugGAN noticeably diminishes the gap. This observation highlights that our DifAugGAN effectively achieves enhanced calibration for the LDL model.}
\vspace{-10pt}
  \label{fig:hist}
\end{figure}

In this paper, we aim to analyze and solve the potential cause of such distortions in the GAN training procedure that prevent it from achieving high-quality SR images. We conjecture that the cause of such distortions is due to the discriminator's inability to send useful signals to the generator to learn good-quality images. To elaborate, during training, we treat our discriminator as a classifier, predicting real or fake images with outputs as the probability of the images being real. As such, a good discriminator should have outputs matching its true probability. For example, an ideal discriminator should be one that has an accuracy of 80\% when its prediction output is 0.8. Interestingly, this coincides with the definition of calibration in the field of conventional classifiers~\cite{guo2017calibration,bai2021don,ferianc2023impact}. To analyze such behaviors, we resort to measuring the expected calibration error (ECE)~\cite{guo2017calibration} in GAN-based SR methods. Significantly, we've noted a pronounced level of calibration in the discriminators throughout the training process, as depicted in Fig.~\ref{fig:hist}. Those poor calibration model is either over-confidence or under-confidence which will amplify or weaken the learning gradient. Consequently, these discriminators are unable to provide the proper gradient information that is crucial for training the generators.

Inspired by the calibration field's techniques that incorporate Gaussian noise into various network components during training to enhance neural network generalization and robustness~\cite{blundell2015weight,neelakantan2015adding,alemi2016deep,guo2017calibration,camuto2020explicit,ash2020warm,ferianc2023impact}, we leverage similar data augmentation strategies to address the calibration issue within discriminators in GAN-based SR models. Differing from prior endeavors, we introduce Gaussian noise into the discriminator's inputs using a diffusion-style approach derived from diffusion models~\cite{ho2020denoising}. Through experimentation, we confirm that this calibration procedure not only yields better-calibrated discriminators, as depicted in Fig.~\ref{fig:hist}, but also effectively resolves distortions inherent in GAN-based approaches, achieving the desired outcomes.
In summary, our contributions are as follows:
\vspace{2pt}
\begin{itemize}
 \item  We identify that discriminators in GAN-based SR methods, like conventional classifiers, are poorly calibrated.
 \vspace{2pt}
 \item  We propose a simple but non-trivial method for calibration of the discriminator during training motivated by the successes of data augmentation schemes in the field to achieve good calibration.
 \vspace{2pt}
 \item  We show through extensive experiments that such an approach can be easily stacked with existing GAN-based SR approaches to improve calibration and thus improve SR performance.
 \item Extensive experiments on synthetic and real-world SISR tasks demonstrate 
demonstrates our DifAugGAN performs superior improvements against the state-of-the-arts both quantitatively and qualitatively.
 
\end{itemize}

\label{sec:intro}

\section{Related Work}
\label{sec:Related_Work}
In this section, we categorize current SISR methods into two main types: PSNR-oriented and GAN-driven approaches. In addition, we also introduce the development of technologies for obtaining well-calibrated neural networks. 

\subsection{PSNR-oriented SR methods} To establish the mapping between HR and LR images, lots of CNN-based methods have emerged~\cite{ledig2017photo,zhang2018image,ma2019matrix,cai2019toward,lyn2020multi}. These methods focus on designing novel architectures by adopting different network modules, such as residual blocks~\cite{lim2017enhanced}, attention blocks~\cite{niu2023gran}, non-local blocks~\cite{mei2021image}, and transformer layers~\cite{liang2021swinir,lu2022transformer}. For example,~\cite{ledig2017photo} employs the ResNet architecture from~\cite{he2016deep} and solves the time and memory issues with good performance. Then~\cite{lim2017enhanced} further optimizes it by analyzing and removing unnecessary modules to simplify the network architecture and produce better results. After them, RCAN~\cite{zhang2018image} and MCAN~\cite{ma2019matrix} adopt the attention mechanism~\cite{woo2018cbam} and design new residual dense networks. Then MLRN ~\cite{lyn2020multi} and BSRT~\cite{li2022blueprint} proposed multi-scale fusion or internal and external features fusion architecture to solve the problem that the existing SISR could not make full use of the characteristic information of the middle network layer and internal features. In addition, SwinIR~\cite{liang2021swinir} and ESRT~\cite{lu2022transformer} apply transformer technology to improve the performance further. 
For optimizing the training process, they prefer to use the MAE or MSE loss (\eg, $L_{1}$ or $L_{2}$) 
to optimize the architectures, which often leads to over-smooth results because the above losses provide a straightforward learning objective and optimize for the popular PSNR (peak signal-to-noise-ratio) metric~\cite{blau2018perception,freirich2021theory,whang2022deblurring}.

\subsection{GAN-driven SR methods}
With generative adversarial
networks (GANs) of all types exhibiting high-quality samples in a wide variety of data modalities, approaches based on the deep generative model have become one of the mainstream for image super-resolution~\cite{wang2018esrgan,soh2019natural,tian2022generative}, which have shown convincing image generation ability. SRGAN~\cite{ledig2017photo} is the first GAN-based SISR method. It adopts the GAN technology to push the generator to produce results with better Visual effects. Compared with SRGAN,   ESRAGN~\cite{wang2018esrgan} trains the discriminator to predict the authenticity of the generated image instead of predicting if the generated image is valid. 
USRGAN~\cite{zhang2020deep} proposes to train an unfolding network, which integrates the merits of traditional model-based methods and CNN-based ones. Then SPSR~\cite{ma2020structure} proposes a structure-preserving method to generate perceptual-pleasant details, which achieves leading performance on synthetic data. BSRGAN~\cite{zhang2021designing} and Real-ESRGAN ~\cite{wang2022realesrgan} extend GAN-based models to real-world applications and obtain promising results, demonstrating their immense potential to restore textures for real-world images. However, GAN-based methods have an obvious drawback in that they optimize the whole training process by the adversarial losses, which often inevitably introduce undesirable artifacts, leading to large distortion, limiting GAN-based SR methods' further application~\cite{lugmayr2021ntire,whang2022deblurring,liang2022details,park2023content,xie2023desra}.

\subsection{Calibration} 
Modern neural networks are known to lack calibration~\cite{neelakantan2015adding,guo2017calibration,bai2021don}. The calibration of a neural network shows that the confidence of the model matches the true probability~\cite{ash2020warm,zhang2022and,ferianc2023impact}.  Moreover, the notion of miscalibration typically involves the disparity between confidence and accuracy. This discrepancy in modern neural networks prompts the use of metrics such as Expected Calibration Error (ECE)~\cite{naeini2015obtaining,guo2017calibration}, which provides a scalar summary of calibration that complements reliability diagrams, allowing for a more convenient and comprehensive assessment of model calibration. One notion of miscalibration is the difference in expectation between confidence and accuracy,\ie,

\begin{equation}
    {\mathbb{E}_k} = \left [ \left | \mathbb{P}\left ( \hat{Y} =Y| \hat{P}=p\right )-p  \right |  \right ] .
    \label{eq:ece}
\end{equation}

The reliability diagrams, depicted at the bottom of Figure~\ref{fig:hist}, serve as a common method to illustrate the calibration of a model prediction~\cite{naeini2015obtaining}. To compute the Expected Calibration Error (ECE), a binary classifier (e.g., discriminator) prediction denoted as $f(x)$ is utilized. Here, the prediction confidence for True is represented by $\sigma(f(x))$, while for False it is $1 - \sigma(f(x))$. To obtain the model's predicted label, $y$, the argmax function is applied to determine $y = \arg\max {\sigma(f(x)), 1 - \sigma(f(x))}$. The process involves equally dividing the prediction confidence into $M$ bins of equal spacing. A weighted average of the accuracy/confidence difference within these bins is then computed. Consequently, the Expected Calibration Error~\cite{naeini2015obtaining}, or ECE, is calculated as:
\begin{equation}
\mathrm{ECE} = \sum_{m=1}^{M} \frac{B_{m}}{n} \left | \mathrm{acc}(B_{m}) - \mathrm{conf}(B_{m})\right | ,
\label{eq:ece_bin}
\end{equation}
where $n$ is the total number of samples, while $B_m$ denotes the samples falling within the $m$th bin. The disparity between $\mathrm{acc}$ and $\mathrm{conf}$ within a specific bin represents the calibration gap. For further insights into this metric, refer to~\cite{naeini2015obtaining,guo2017calibration} for comprehensive analyses.

To improve the calibration of the model during the training, many kinds of methods are proposed. One of the simple and effective ways is data augmentation, e.g. MixUp~\cite{zhang2018mixup} and ODS~\cite{tashiro2020diversity}. Those techniques can be applied in a model architecture and task-agnostic way. MixUp regularizes the neural network to favor simple linear behavior in between training examples~\cite{zhang2018mixup}, which has been proven to improve calibration in~\cite{zhang2022and}. ODS proposes a novel sampling strategy that attempts to maximize diversity in the target model outputs among the generated samples, which has been widely used in adversarial attack tasks to
improve the robustness of neural
networks (NN).  Furthermore, ~\cite{blundell2015weight,alemi2016deep,camuto2020explicit} try to add Gaussian noise to the activation of NN, while uniform noise injection adds uniform noise~\cite{neelakantan2015adding} has shown adding Gaussian noise to the gradients
during training can improve the generalization of neural network. 
~\cite{ash2020warm} adding Gaussian noise via shrinking and perturbing
weights of neural networks to improve retraining generalization. In addition, ~\cite{ferianc2023impact} has shown
adding Gaussian noise during training is effective in improving calibration. These methods give us much inspiration that introducing noise to various components of the network, including the input, targets, activation, gradients, and the model itself during the training helps to explore more effective neural network models.

\begin{figure*}[!htbp]
  \centering
  \includegraphics[width=0.95 \linewidth]{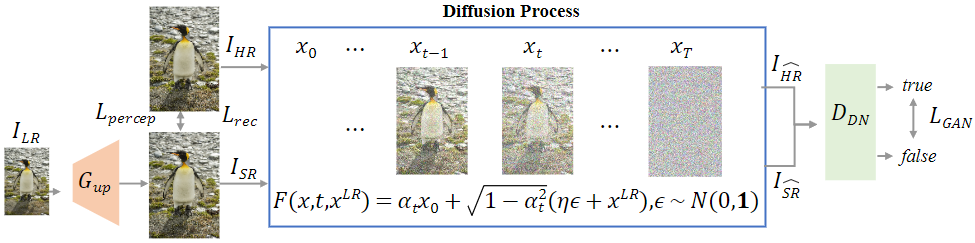}
   \caption{Overall framework of DifAugGAN. It consists of three parts. An upsampling generator to generate $I_{SR}$ for the input $I_{LR}$. The diffusion process to add noise to the inputs of the discriminator. The discriminator to recognize its input true or false.}
  \label{fig:overview}
\end{figure*}

\section{Perliminaries}
To avoid ambiguity, we start by presenting the problem of our interest,~\ie, GAN-based SR, and how it is performed in current literatures~\cite{liang2022details,xie2023desra}.

\textbf{GAN-based super resolution} target to reconstruct a high perceptual quality SR image $x^{SR}\in \mathcal{R}^{H\times W \times 3}$ from a given low-resolution image $x^{LR}$. To accomplish this goal, a generator model $G_\theta$ is usually trained with the following three sorts of losses~\cite{wang2018esrgan, wang2022realesrgan, zhang2021designing}:
\begin{equation}
\label{eq:g_train}
    \mathcal{L}_G=\mathcal{L}_{pixel} +\lambda_1 \mathcal{L}_{percep} +\lambda_2 \mathcal{L}_{adv},
\end{equation}
where $\mathcal{L}_{pixel}$ denotes a reconstruction loss such as $l_1$ or $l_2$ distance.  $\mathcal{L}_{percep}$ represents a perceptual loss~\cite{zhang2018unreasonable} measure the feature distance and $\mathcal{L}_{adv}$ is the adversarial loss. Specifically, the adversarial loss involves the utilization of a discriminator network $D_\theta$ that classifies a high-resolution image $x^{HR}$ as TRUE and an image generated from generator $\hat{y}$ as FALSE:
\begin{equation}
    \left\{ 
    \begin{aligned}
        D_\theta(x)\rightarrow 1,&~~ \text{if $x=x^{HR}$} \\
        D_\theta(x)\rightarrow 0,&~~ \text{if $x=x^{LR}$} \\ 
    \end{aligned}
    \right. .
\end{equation}
The training objective for the discriminator then is usually defined as follows:
\begin{equation}
    \mathcal{L}_D=-\log(1-\sigma(D(x^{HR}))) - \log(\sigma(D(\hat{y}))),
\end{equation}
where $\sigma(\cdot)$ is the sigmoid function. Furthermore, certain studies \cite{wang2018esrgan,liang2022details} have explored the utilization of relativistic GAN as their GAN loss, which can be considered an extended case or a variant of the normal GAN loss. However, a comprehensive discussion of this aspect is deferred to future investigations.

\textbf{Two phases training.} To get the best performance, GAN-based SR typically follows a two-step training process. Initially, the generator is trained using $\mathcal{L}_{pixel}$ to optimize for Peak Signal-to-Noise Ratio (PSNR), thus yielding a model tend to output over-smoothed image~\cite{wang2018esrgan}. Subsequently, the generator undergoes further fine-tuning by incorporating $\mathcal{L}_G$ and $\mathcal{L}_D$ to encourage the model to generate realistic images. However, the adversarial optimization in the second stage is unstable and  makes the preceding SR model generate unpleasant and undesirable artifacts, leading to large distortion.~\cite{liang2022details,xie2023desra}. We attribute the cause of such distortions to the discriminator's poor calibration makes its prediction either over confidence or under confidence. Thus the discriminator is unable to send useful signals to the generator to learn good-quality images. In this work, we focus on improving the second stage of GAN-based SR training, and the first stage are keeping the same with existing works~\cite{wang2018esrgan,wang2022realesrgan}.

\section{Method}
Building upon existing studies highlighting noise injection as a successful method for improving calibration~\cite{camuto2020explicit,ash2020warm,ferianc2023impact}, we introduce a novel and dynamic diffusion-style noise augmentation strategy aimed at addressing the prevalent issue of subpar calibration in contemporary GAN-based SR training. Visualized in Fig.~\ref{fig:overview}, our strategy encompasses three essential components: 1) an up-sampling generator denoted as $G_{up}$, 2) a diffusion processor characterized by $\mathcal{F}$, and 3) a discriminator $D_{DN}$. In the following content, we will introduce the whole workflow in detail.

\textbf{Diffusion process on discriminator.} The core idea of our method is adding Gaussian noise on the input of the discriminator which we call it discriminator diffusion process. $\mathcal{F}_{dis}$.
In normal image generative model~\cite{ho2020denoising} the diffusion process $\mathcal{F}$ is described as gradually adding Gaussian noise into a data sample according to a noise schedule $\{\alpha_1,...,\alpha_T\}$ with a total of $T$ steps. Given a data sample $x_0$, we define the processor at arbitrary time step $t$ as follows:
\begin{equation}
    \mathcal{F}(x_0, t) = \alpha_t \bx_0 + \sqrt{1-\alpha_t^2}\epsilon,\epsilon\sim\mathcal{N}(0,I).
\end{equation}
The noise schedule $\alpha_t$ is usually a monotonically decreasing function where $\alpha_0=1, \alpha_T=0$. Thus, given a training dataset with a high-resolution image and a low-resolution pair $(x,y)\sim\mathcal{X}$. We can formulate our diffusion processor $\mathcal{F}$ which takes input of a generator output or the corresponding low-resolution image. Specifically, different from the diffusion process in the stand DDPM~\cite{ho2020denoising}, we find that making the mean value of the Gaussian noise added in each step has the mean value of a bicubic-upsampled low-resolution image instead of zero mean helps the model get better performance. Our discriminator diffusion process is therefore formulated as follows:
\begin{equation}
    \mathcal{F}(x, t, x^{LR}) = \alpha_t \bx_0 + \sqrt{1-\alpha_t^2}(\eta\epsilon+x^{LR}),\epsilon\sim\mathcal{N}(0,I),
\end{equation}
where $\eta$ downscaling the variance of the noise. Then our final discriminator training can be formed as follows:
\begin{equation}
    \mathcal{L}_{DifAug}=-\log(1-\sigma(D(\mathcal{F}(x^{HR})))) - \log(\sigma(D(\mathcal{F}(\hat{y})))).
\end{equation}
Correspondingly, we can obtain a  new adversarial loss for the generator can be written as follows:
\begin{equation}
    \mathcal{L}_{adv} = -\log(1-\sigma(D(\mathcal{F}(\hat{y})))).
\end{equation}

As for the noise schedule for the diffusion process, in this paper, we mainly consider the linear schedule proposed in DDPM~\cite{ho2020denoising}:
\begin{equation}
    \alpha_t=\exp{[-\frac{1}{4}t^2 (\beta_{max}-\beta_{min})-\frac{1}{2}t\beta_{min}] },
\end{equation}
where $\beta_{min}=0.1$ and $\beta_{max}=20$ in all of our experiment.
After training with the new loss adversarial loss $\mathcal{L}_{adv}$, the calibration of the discriminator will be optimized and then stabilize the whole adversarial training, alleviating the distortions resulting from GANs’ unstable optimization. 

\begin{figure*}[!h]
\centering
  \includegraphics[width=0.95 \linewidth]{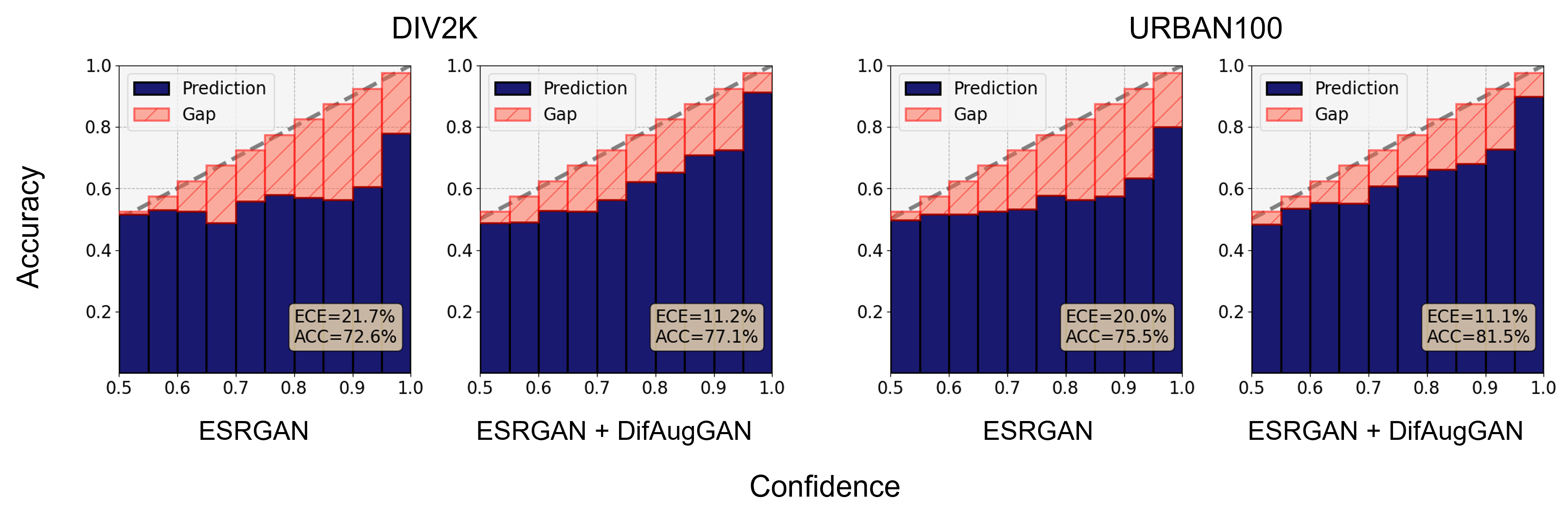}
  \caption{Reliability graph for ESRGAN model and training with our DifAugGAN on DIV2K100 and URBAN100 testing dataset. Both models are trained on the DIV2K dataset. It shows the alignment of the model prediction with the gap between the ideal calibration model.}
  \label{fig:reliability_div2k}
\end{figure*}

\begin{table}[!htbp]
\centering
\caption{Comparision different method in terms of ECE~\cite{dong2015image,guo2017calibration} and accuracy of the discriminator on the real-world scenario.}
\label{tab:ece_acc}
\resizebox{0.9\hsize}{!}{
\begin{tabular}{ c|c|lc}
\toprule
\hline
Dataset & Method & ECE $\downarrow$ (\%) & ACC $\uparrow$ (\%) \\ \hline
\multirow{4}{*}{DIV2K}   & RealESRGAN              & 8.8    &  97.0 \\ 
                         & +\textbf{DifAugGAN}        & 7.4 (\textbf{-1.4}) & 98.8 \\ \cline{2-4} 
                         & RealESRGAN + LDL        & 9.6 &  95.2   \\
                         & +\textbf{DifAugGAN}        & 5.6 (\textbf{-4.0}) &  97.7 \\ \hline
\multirow{4}{*}{Urban100} & RealESRGAN              & 6.9 & 97.5 \\ 
                         & +\textbf{DifAugGAN}        & 2.8 (\textbf{-4.2}) &   97.5 \\ \cline{2-4} 
                         & RealESRGAN + LDL        & 8.1  &  95.3  \\ 
                         & +\textbf{DifAugGAN}        &  3.1 (\textbf{-5.0}) & 97.4 \\ \hline
                         \bottomrule
\end{tabular}}
\end{table}

\section{Experiments}

\begin{table*}[hbt!]
\centering
\caption{Quantitative comparison between the proposed DifAugGAN and SOTA GAN-SR methods. \textbf{bold +DifAugGAN} represents our model, which takes the method in its before column as its baseline. From this table, we can find whether it is ESRGAN or LDL, with the addition of our DifAugGAN, the performance has been significantly improved on different training sets. The best and second-best results are highlighted
in \textcolor{red}{red} and \textcolor{blue}{blue}. $\uparrow$ and $\downarrow$ mean that the larger or smaller score is better, respectively.}
\label{tab:benchmark}
\resizebox{1\hsize}{!}{
\begin{tabular}{cc|cccc|cccccccc}
\toprule
\hline
\multicolumn{2}{c|}{Method}   & ESRGAN & \textbf{+DifAugGAN} & LDL  & \textbf{+DifAugGAN}   & ESRGAN &\textbf{+DifAugGAN} & USRGAN & SPSR   & LDL  & \textbf{+DifAugGAN} & SwinIR+LDL & \textbf{+DifAugGAN} \\
\multicolumn{1}{c}{Metric}   & Dataset & \multicolumn{4}{c|}{DIV2K}   & \multicolumn{8}{c}{DF2K} 
\\ \hline
\multicolumn{1}{c}{}   & Set5   & 30.095 &  30.769     & 30.985 & \color{red}{31.416}    &  30.438 &  31.430     &  30.911 &  30.397 &  31.033 &  \color{blue}{31.475}   &  31.028     & {\color{red}{31.733}}        \\ 

\multicolumn{1}{c}{}                          & Set14   &  26.738 &  27.358     &  27.491 &  \color{red}{27.919}    &  26.594 &  27.872     &  27.405 &  26.860 &  27.228 &  \color{blue}{27.883}   &  27.526     &{\color{red}{ 28.183}}        \\ 

\multicolumn{1}{c}{}    & Urban100   &  24.668 &  25.136     &  \color{blue}{25.498} &  \color{red}{25.842}                                 &  24.365 &  25.923     &  24.891 &  24.804 &  25.458 &  25.923   &  \color{blue}{26.231}     & \color{red}{26.543 }       \\

\multicolumn{1}{c}{}   & DIV2K(val)   &  28.132 &  28.787     &  28.951 &  \color{red}{29.373}        &  28.175 &  \color{blue}{29.456}    &  28.787 &  28.182 &  28.818 &  29.437   &  29.117     & \color{red}{29.820}        \\ 

\multicolumn{1}{c}{\multirow{-5}{*}{PSNR$\uparrow$}}    & General100 &  29.489 &  30.023     &  \color{blue}{30.232}                                & 30.206  &  29.426 &  30.775     &  30.002 &  29.423 &  30.289 &  \color{blue}{30.783}   &  30.441     & \color{red}{31.014 }       \\ \hline

\multicolumn{1}{c}{}     & Set5    &  0.8415 &  0.8598     &  \color{blue}{0.8626} &  \color{red}{0.8730}                                 &0.  8513   &  \color{blue}{0.8724}    &  0.8658 &  0.8443 &  0.8611 &  0.8741   &  0.8611     & \color{red}{0.8782}        \\

\multicolumn{1}{c}{}        & Set14   &  0.7204 &  0.7442     &  0.7476 &  \color{red}{0.7603}        &  0.7168 &  \color{blue}{0.7587}     &  0.7486 &  0.7254 &  0.7358 &  0.7585   &  0.7478    & \color{red}{0.7706}        \\ 

\multicolumn{1}{c}{}                          & Urban100   &  0.7405 &  0.7551     &  \color{blue}{0.7673} &  \color{red}{0.7770}                                 &  0.7364 &  0.7802    &  0.7503 &  0.7474 &  0.7661 &  0.7795   &  \color{blue}{0.7918}    & \color{red}{0.7998}        \\

\multicolumn{1}{c}{}                          & DIV2K(val)   &  0.7722 &  0.7903     &  \color{blue}{0.7951} &  \color{red}{0.8049}                                 &  0.7759 &  \color{blue}{0.8067}    &  0.7941 &  0.7720 &  0.7897 &  0.8060   &  0.8011     & \color{red}{0.8193}        \\

\multicolumn{1}{c}{\multirow{-5}{*}{SSIM$\uparrow$}}    & General100 &  0.8103 &  0.8267     &  \color{blue}{0.8277} &  \color{red}{0.8367}                                &  0.8095 &  \color{blue}{0.8406}     &  0.8241 &  0.8091 &  0.8280 &  \color{blue}{0.8406}  &  0.8347     & \color{red}{0.8482}        \\  \hline

\multicolumn{1}{c}{}                          & Set5    &  \color{blue}{0.0690} &  0.0708     &  \color{red}{0.0670} &  0.0763   &  0.0758 &  0.0795     &  0.0795 &  \color{red}{0.0647} &  0.0690 &  0.0809   &  0.0655    & \color{blue}{0.0654}        \\ 

\multicolumn{1}{c}{}                          & Set14   &  \color{blue}{0.1236} &  0.1298     &  \color{red}{0.1207} &  0.1315   &  0.1241 &  0.1188    &  0.1347 &  0.1207 &  0.1132 &  0.1189   &  \color{red}{0.1091}    & \color{blue}{0.1117}        \\ 

\multicolumn{1}{c}{}                          & Urban100   &  0.1231 &  0.1238    &  \color{red}{0.1096}   & \color{blue}{0.1167}  &  0.1229 &  0.1152     &  0.1330 &  0.1184 &  0.1084 &  0.1156  &  \color{red}{0.1021}    &\color{blue}{ 0.1041}        \\

\multicolumn{1}{c}{}                          & DIV2K(val)   &  0.1137 &  0.1117    &  \color{red}{0.1011} &  \color{blue}{0.1085}    &  0.1154 &  0.1074    &  0.1325 &  0.1099 &  0.0999 &  0.1074   &  \color{red}{0.0944}     & 0.0969        \\  

\multicolumn{1}{c}{\multirow{-5}{*}{LPIPS$\downarrow$}}    & General100 &  0.0876 &  0.0873     &  \color{red}{0.0790} &  \color{blue}{0.0842}    &  0.0879 &  0.0865     &  0.0937 &  0.0862 &  0.0796 &  0.0876   &  0.0740     & \color{red}{0.0735 }       \\ \hline

\multicolumn{1}{c}{}                          & Set5    &  0.0940 &  0.0987    &  \color{red}{0.0917} &  \color{blue}{0.0976}                               &  0.0950 &  0.0990    &  0.1045 &  0.0922 &  0.0919 &  0.1029  &  \color{red}{0.0899}     & \color{blue}{0.0912}        \\ 

\multicolumn{1}{c}{}                          & Set14   &  0.0977 &  0.0996    &  0.0935 &  \color{blue}{0.0968}    &  0.0951 &  0.0924     &  0.0997 &  0.0921 &  \color{red}{0.0866} &  0.0945  &  \color{blue}{0.0869}     & 0.0888        \\ 

\multicolumn{1}{c}{}                          & Urban   &  0.0898 &  0.0909    &  \color{red}{0.0822} &  \color{blue}{0.0869}      &  0.0880 &  0.0859     &  0.0975 &  0.0849 &  0.0793 &  0.0861  &  \color{red}{0.0800}    &\color{red}{ }0.0800        \\ 

\multicolumn{1}{c}{}                          & DIV2K(val)   &  0.0574 &  0.0576     &  \color{red}{0.0528} &  0.0585  &  0.0594 &  0.0588    &  0.0645 &  0.0546 &  0.0526 &  0.0588   &  \color{red}{0.0507}     & \color{blue}{0.0520}        \\ 

\multicolumn{1}{c}{\multirow{-5}{*}{DIST$\downarrow$}}    & General100 &  0.0896 &  0.0892     &  \color{red}{0.0827} &  0.0854   &  0.0874 &  0.0855     &  0.0931 &  0.0885 &  0.0802 &  0.0855   &  \color{blue}{0.0794}     & \color{red}{0.0782}        \\ \hline

\hline
\end{tabular}
}
\end{table*}

\indent In this section, we conduct extensive experiments to verify the effectiveness of the proposed DifAugGAN. We first introduce the experiment setups and then conduct analyses of how  DifAugGAN helps to achieve well-calibrated discriminators for GAN-based SR methods on both synthetic and real-world datasets. Finally, we report the qualitative and quantitative results obtained by comparing  DifAugGAN with  SOTA methods.
\subsection{Experiment setup}
\textbf{Implementation details.} For the synthetic experiments, we adopt the DIV2K~\cite{agustsson2017ntire} including 800 image pairs and DF2K~\cite{lim2017enhanced,timofte2017ntire} including 
3450 image pairs with the scaling factor of 4$\times$  as the training dataset respectively. We employ benchmark datasets including Set5~\cite{bevilacqua2012low}, Set14~\cite{zeyde2012single}, Urban100~\cite{huang2015single}, DIV2K 
validation~\cite{agustsson2017ntire}, and General100~\cite{dong2016accelerating} for validation. For real-world image super-resolution, we train our model on DF2K+OST~\cite{wang2022realesrgan} with the scaling factor of 4$\times$ and adopt  
DIVKRK~\cite{bell2019blind}, NTIRE20~\cite{lugmayr2020ntire}, RealSR(V3)~\cite{cai2019toward},
RealSRset~\cite{zhang2021designing}, and RealSR~\cite{ji2020real} as the validation dataset. All the experiments are run with 4 NVIDIA GTX A100 GPUs and the batch size is 16 per GPU for synthetic experiments and 12 per GPU for real word experiments. Following~\cite{liang2022details}, we use Adam optimizer with learning rate 1e-4. The total training iterations are $300K$. For real-world experiments, we use the same setting with Real-ESRGAN~\cite{wang2022realesrgan}. For all experiments, we set the noise downscaling factor $\eta=0.05$ and the total time step $T=1000$.

\noindent\textbf{Compared methods.} 
We employ ESRGAN~\cite{wang2018esrgan} and LDL~\cite{liang2022details}
as our baselines to validate the effectiveness of the proposed DifAugGAN, resulting in our models ESRGAN+DifAugGAN and LDL+DifAugGAN. In addition, we also use  SwiniR~\cite{liang2021swinir} as the generator model to take the place of the RRDB-based generator model in  LDL, resulting in a new baseline  SwinIR+LDL for our DifAugGAN. In addition, we also compare our model with USRGAN~\cite{zhang2020deep} and SPSR~\cite{ma2020structure}.
We further validate DifAugGAN for conducting real-world SISR tasks by employing DifAugGAN to Real-ESRGAN~\cite{wang2022realesrgan} and LDL~\cite{liang2022details}, resulting in Real-ESRGAN+DifAugGAN and LDL+DifAugGAN.  Then we compare them with BSRGAN~\cite{zhang2021designing}, Real-ESRGAN, and LDL models.

\subsection{ Calibration Improvements Analysis}
The introduced DifAugGAN employs a diffusion-style data augmentation scheme which can enhance the calibration of the discriminator during training. This strategy enables the discriminator to provide more accurate and informative feedback to the generator, fostering the production of high-quality, realistic images. To evaluate the effectiveness of our proposed method for calibration, we mainly apply ECE metric to measure the confidence and accuracy.

\noindent\textbf{Calibration on Synthetic Datasets.} Our analysis, as depicted in Fig.~\ref{fig:reliability_div2k}, demonstrates a remarkable reduction in ECE when DifAugGAN is integrated on top of the ESRGAN baseline. Specifically, the ECE witnesses a halving on both the DIV2K validation dataset and the URBAN100 dataset, using 1000 random crops from high-resolution and super-resolved images. Concurrently, there is an improvement in discriminator accuracy. The improved calibration, denoted by lower ECE values, enables the discriminator to furnish more accurate and informative feedback to the generator. This calibration refinement, coupled with a well-calibrated discriminator, augments the generation of high-quality, photorealistic images. Notably, the ESRGAN model used in this analysis is trained by us, employing normal GAN loss as opposed to the relativistic GAN loss stipulated in the original paper.

\noindent\textbf{Calibration on Real-World Dataset.} Assessing our method in a real-world scenario involves comparison within setups like RealESRGAN and RealESRGAN+LDL. The testing dataset, DIV2K100 and URBAN100, undergoes the same processing strategy as RealESRGAN when fed into the discriminator model. The comparison results illustrated in Tab.~\ref{tab:ece_acc} reveal a consistent and significant decrease in ECE across both DIV2K and Urban100 testing datasets with the incorporation of DifAugGAN compared to baseline methods. This reduction indicates an enhanced alignment between predicted confidence and actual accuracy, validating the effectiveness of DifAugGAN in enhancing calibration. While maintaining or slightly improving discriminator accuracy in comparison to baseline methods, DifAugGAN notably enhances calibration. This enhancement is crucial as it ensures that the confidence estimates provided by the discriminator are more dependable, guiding the generator toward producing higher-quality images.
\vspace{20pt}

\begin{figure*}[!htbp]
  \centering
  \includegraphics[width=1 \linewidth]{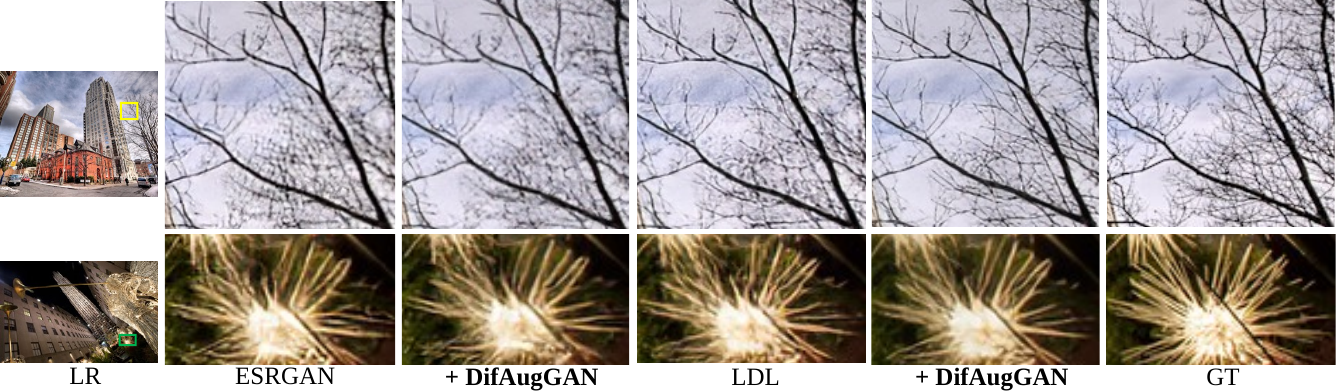}
  \caption{Visual comparison to state-of-the-art GAN-SR methods. LR images are from Urban100 and General100. The~\textbf{bold +DifAugGAN} represents our method, which employs our DifAugGAN to its baseline  in front of it. All methods are trained on DIV2K. More visual comparisons can be found in the supplementary materials. ($\times$4 scale. Zoomed in for a better view.)  }
  \label{fig:DIV2K}
\end{figure*}

\begin{figure*}[!htbp]
  \centering
  \includegraphics[width=1 \linewidth]{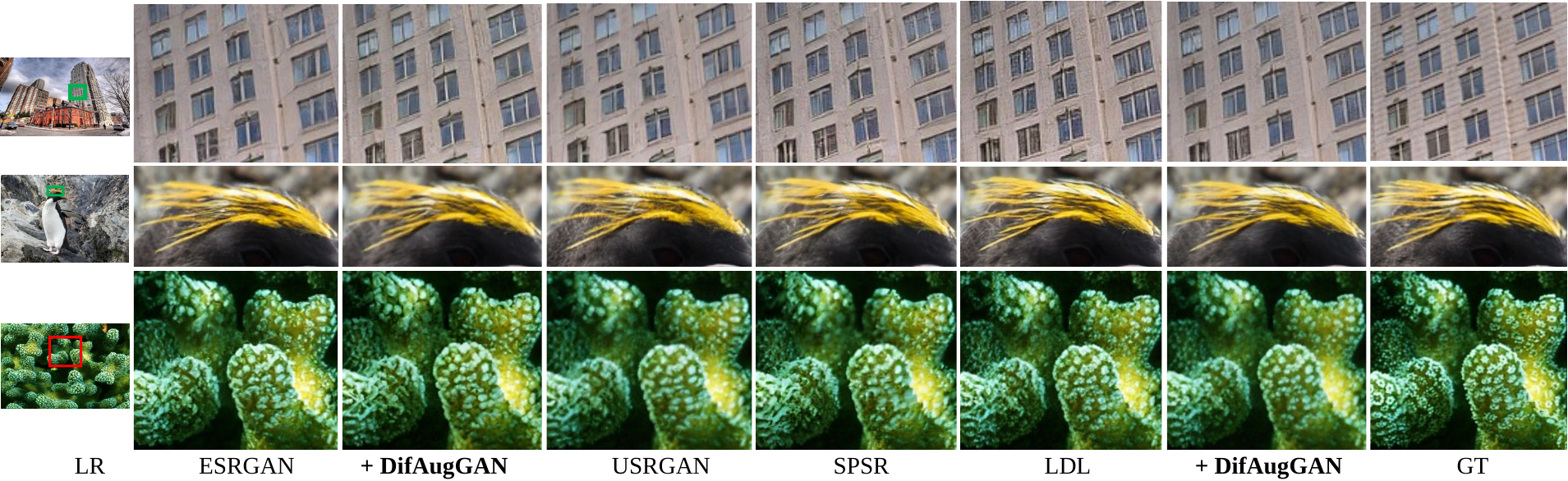}
  \caption{Visual comparison to state-of-the-art GAN-SR methods. LR images are from Urban100, DIV2K100, and General100, respectively. The~\textbf{bold +DifAugGAN} represents our method, which employs our DifAugGAN to its baseline  in front of it. All methods are trained on DF2K. More visual comparisons can be found in the supplementary materials. ($\times$4 scale. Zoomed in for a better view.)  }
  \vspace{-10pt}
  \label{fig:DF2K}
\end{figure*}

\subsection{Validation on synthetic datasets}

\noindent\textbf{Qualitative comparison.} Fig.~\ref{fig:DIV2K} and Fig.~\ref{fig:DF2K} illustrate the visual comparisons among the proposed DifAugGAN and its baselines and other SOTA GAN-based SR methods. Obviously, for the baselines,~\ie ESRGAN+DifAugGAN, LDL+DifAugGAN, after adopting our DifAugGAN, the artifacts in the outputs have been significantly weakened, especially in regions where there is a lot of regional duplication (\eg buildings, windows, and hair.). Furthermore, superfluous details have also been alleviated, such as the small flowers in the green plants.   When compared with BSRGAN and USRGAN, our models still perform superior.
This effectiveness makes DifAugGAN a practical and effective framework to further improve  GAN-based SR methods.

\noindent\textbf{Quantitative comparison.} We first report the performance of the proposed DifAugGAN and the state-of-the-art GAN-based SR methods on synthetic datasets in Tab.~\ref{tab:benchmark}. We trained the methods on both DIV2K and DF2K respectively. The~\textbf{bold +DifAugGAN} represents our model which takes its before line as its baseline. It is obvious that the proposed DifAugGAN can achieve a better trade-off between the distortion quality metrics (PSNR, SSIM) and perceptual quality metrics (LPIPS, DISTS,)
on most benchmarks than the three baselines,~\ie, ESRGAN+DifAugGAN, LDL+DifAugGAN, 
and  SwinIR+LDL+DifAugGAN, though our DifAugGAN models perform a little worse on perceptual quality metrics  LPIPS \textbf{+0.0003 $\sim$ 0.001} and DISTS \textbf{+0.0004 $\sim$ 0.006}  than their baselines.
In addition, among LDL, USRGAN, and SPSR, there also subtle fluctuations exist in their perceptual quality metrics. We attribute these fluctuations to the properties of GANs, which are famous for producing overly realistic image details than ground truth, resulting in good perceptual quality metrics. 
After performing the artifact elimination methods, the artifacts generated by GAN are eliminated, along some of the details generated by GAN will inevitably be eliminated, which will lead to worse perceptual quality metrics.
In terms of distortion quality metrics PSNR and SSIM, our DifAugGAN models obtain obvious improvement with PSNR \textbf{+0.4} $\sim$ \textbf{0.7dB} and SSIM \textbf{+0.01} $\sim$ \textbf{+0.03} over their baselines, successively. When compared with USRGAN and BSRGAN, our models still have overwhelming advantages.

\begin{figure*}[!htbp]
  \centering
  \includegraphics[width=1 \linewidth]{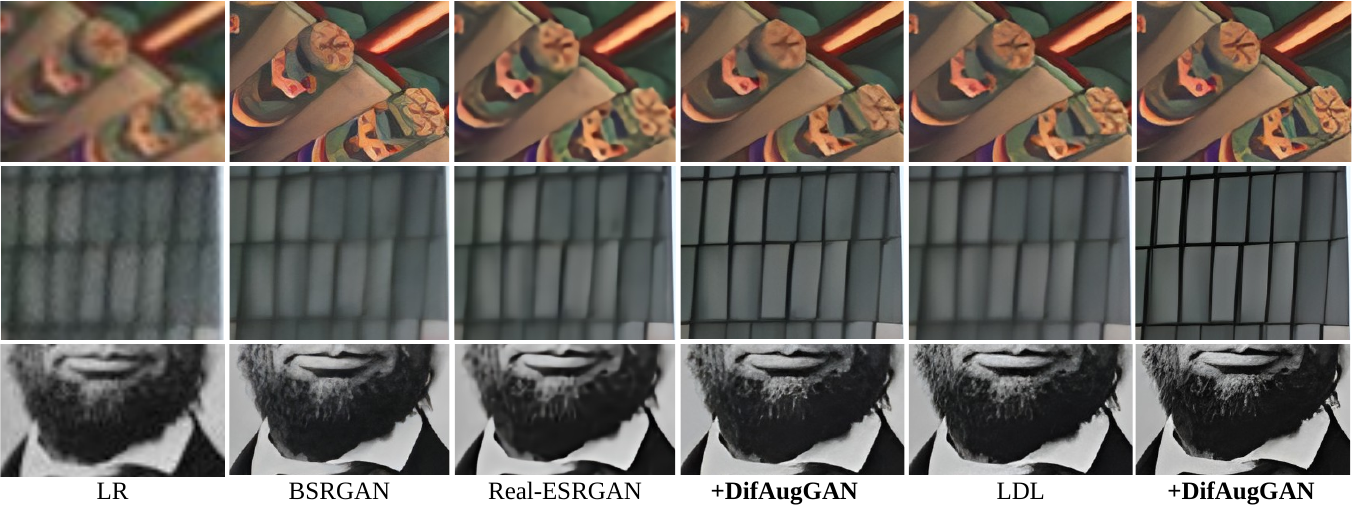}
  \caption{Visual comparison to state-of-the-art GAN-SR methods. LR images are from RealSR~\cite{cai2019toward}, NTIRE20~\cite{lugmayr2020ntire}, RealSet~\cite{zhang2021designing}, respectively. All methods are trained on DF2K+OST. The~\textbf{bold +DifAugGAN} represents our method, which employs our DifAugGAN to its baseline  in front of it. More visual comparisons can be found in the supplementary materials. ($\times$4 scale. Zoomed in for a better view.)  }
  \label{fig:real_2}
\end{figure*}

\subsection{Validation on Real-world datasets}
To demonstrate the generalization of the proposed DifAugGAN, we apply it to conduct experiments on real-world images. We take DF2K+OST as the training set, RealSR~\cite{cai2019toward}, NTIRE20 track2~\cite{lugmayr2020ntire}, RealSet~\cite{zhang2021designing}, DIVKRK~\cite{bell2019blind}, and  RealSR~\cite{ji2020real}.  We take Real-ESRGAN~\cite{wang2022realesrgan}, LDL~\cite{liang2022details} as our baseline. In addition, we also adopt BSRGAN~\cite{zhang2021designing} as a comparative method. 
Since there is no ground truth for real-world images, we give qualitative results to compare the proposed DifAugGAN with the above methods.

\noindent\textbf{Qualitative comparison.} 
As shown in Fig.\ref{fig:real_2}, It is obvious that our DifAugGAN can strongly remove the artifacts generated from Real-ESRGAN and LDL for the patterned building structures in the first row, where BSRGAN produces over-smooth results. For the areas with regular patterns in the second row, our DifAugGAN encourages Real-ESRGAN and LDL to generate sharper details. For the beard with dense textures in the third row, our DifAugGAN pushes its baselines to generate more realistic details.

\subsection{Ablation study}
In order to better understand our method, we explore the effect of adding noise. We show the result when we do not add the low-resolution image into the noise. As shown in Tab.~\ref{tab:addlr}, we can see that adding the noise into Gaussian can help the discriminator learn better. Both the fidelity-oriented metric (PSNR) and perceptual quality metric (LPIPS) boost the performance after adding the low-resolution image. Thus when generating the noise for our method, we add an upsampled low-resolution image as the mean value.
\begin{table}[!ht]
    \centering
    \caption{Results when adding up-sampled low-resolution image into the Gaussian noise.}
     \label{tab:addlr}
    \resizebox{0.7\hsize}{!}{
    \begin{tabular}{c|c|c|c}
    \toprule
    \hline
        Dataset & Noise & PSNR $\uparrow$ & LPIPS $\downarrow$ \\ \hline
        Set14 & Gaussian & 27.30 & 0.1302 \\ 
        ~ & +LR & 27.35 & 0.1096 \\ \hline
        DIV2K & Gaussian & 28.70 & 0.1135 \\ 
        ~ & +LR & 28.78 & 0.1011 \\\hline
        General100 & Gaussian & 29.97 & 0.0875 \\ 
        ~ & +LR & 30.02 & 0.0790 \\
        \hline
    \bottomrule
    \end{tabular}}
    \vspace{-10pt}
\end{table}

\subsection{Limitations} 
While DifAugGAN outperforms existing methods, the specific mechanisms through which the diffusion-style augmentation improves discriminator calibration might require further investigation and validation. More in-depth analysis could provide better insights into the underlying reasons for its success and potential limitations. Also, the effectiveness of the diffusion-style data augmentation in DifAugGAN could heavily rely on the proper calibration of diffusion parameters. Finding optimal parameter values might be challenging, and the performance of DifAugGAN could be sensitive to these choices. Addressing these limitations could provide a clearer understanding of the scope and applicability of the proposed DifAugGAN approach in GAN-based image super-resolution methods.

\section{Conclusion}

In this paper, we propose a diffusion-style data augmentation scheme to solve the issue of the calibration of discriminators in GAN-based SR models, known as DifAugGAN, to alleviate the large distortion  result from GAN-based  SR models. It involves introducing the diffusion process in diffusion models for calibration of the discriminator during training motivated by the successes of data augmentation schemes in the field to achieve good calibration. Our DifAugGAN can be a Plug-and-Play strategy for current GAN-based SISR methods to improve the calibration of the discriminator and thus improve SR performance. Extensive experiments on synthetic and real-world SISR datasets demonstrate that our DifAugGAN outperforms the existing GAN-SR methods both quantitatively and qualitatively.

{
    \small
    \bibliographystyle{ieeenat_fullname}
    \bibliography{main}

\begin{thebibliography}{58}
\providecommand{\natexlab}[1]{#1}
\providecommand{\url}[1]{\texttt{#1}}
\expandafter\ifx\csname urlstyle\endcsname\relax
  \providecommand{\doi}[1]{doi: #1}\else
  \providecommand{\doi}{doi: \begingroup \urlstyle{rm}\Url}\fi

\bibitem[Agustsson and Timofte(2017)]{agustsson2017ntire}
Eirikur Agustsson and Radu Timofte.
\newblock Ntire 2017 challenge on single image super-resolution: Dataset and study.
\newblock In \emph{CVPRW}, 2017.

\bibitem[Alemi et~al.(2017)Alemi, Fischer, Dillon, and Murphy]{alemi2016deep}
Alexander~A Alemi, Ian Fischer, Joshua~V Dillon, and Kevin Murphy.
\newblock Deep variational information bottleneck.
\newblock In \emph{ICLR}, 2017.

\bibitem[Ash and Adams(2020)]{ash2020warm}
Jordan Ash and Ryan~P Adams.
\newblock On warm-starting neural network training.
\newblock \emph{NeurIPS}, 2020.

\bibitem[Bai et~al.(2021)Bai, Mei, Wang, and Xiong]{bai2021don}
Yu Bai, Song Mei, Huan Wang, and Caiming Xiong.
\newblock Don’t just blame over-parametrization for over-confidence: Theoretical analysis of calibration in binary classification.
\newblock In \emph{ICML}, 2021.

\bibitem[Bell-Kligler et~al.(2019)Bell-Kligler, Shocher, and Irani]{bell2019blind}
Sefi Bell-Kligler, Assaf Shocher, and Michal Irani.
\newblock Blind super-resolution kernel estimation using an internal-gan.
\newblock \emph{NeurIPS}, 2019.

\bibitem[Bevilacqua et~al.(2012)Bevilacqua, Roumy, Guillemot, and Alberi-Morel]{bevilacqua2012low}
Marco Bevilacqua, Aline Roumy, Christine Guillemot, and Marie~Line Alberi-Morel.
\newblock Low-complexity single-image super-resolution based on nonnegative neighbor embedding.
\newblock 2012.

\bibitem[Blau and Michaeli(2018)]{blau2018perception}
Yochai Blau and Tomer Michaeli.
\newblock The perception-distortion tradeoff.
\newblock In \emph{CVPR}, 2018.

\bibitem[Blundell et~al.(2015)Blundell, Cornebise, Kavukcuoglu, and Wierstra]{blundell2015weight}
Charles Blundell, Julien Cornebise, Koray Kavukcuoglu, and Daan Wierstra.
\newblock Weight uncertainty in neural network.
\newblock In \emph{ICML}, 2015.

\bibitem[Cai et~al.(2019)Cai, Zeng, Yong, Cao, and Zhang]{cai2019toward}
Jianrui Cai, Hui Zeng, Hongwei Yong, Zisheng Cao, and Lei Zhang.
\newblock Toward real-world single image super-resolution: A new benchmark and a new model.
\newblock In \emph{ICCV}, 2019.

\bibitem[Camuto et~al.(2020)Camuto, Willetts, Simsekli, Roberts, and Holmes]{camuto2020explicit}
Alexander Camuto, Matthew Willetts, Umut Simsekli, Stephen~J Roberts, and Chris~C Holmes.
\newblock Explicit regularisation in gaussian noise injections.
\newblock \emph{NeurIPS}, 2020.

\bibitem[Dong et~al.(2015)Dong, Loy, He, and Tang]{dong2015image}
Chao Dong, Chen~Change Loy, Kaiming He, and Xiaoou Tang.
\newblock Image super-resolution using deep convolutional networks.
\newblock \emph{TPAMI}, 2015.

\bibitem[Dong et~al.(2016)Dong, Loy, and Tang]{dong2016accelerating}
Chao Dong, Chen~Change Loy, and Xiaoou Tang.
\newblock Accelerating the super-resolution convolutional neural network.
\newblock In \emph{ECCV}, 2016.

\bibitem[Ferianc et~al.(2023)Ferianc, Bohdal, Hospedales, and Rodrigues]{ferianc2023impact}
Martin Ferianc, Ondrej Bohdal, Timothy Hospedales, and Miguel Rodrigues.
\newblock Impact of noise on calibration and generalisation of neural networks.
\newblock \emph{ICMLW}, 2023.

\bibitem[Freirich et~al.(2021)Freirich, Michaeli, and Meir]{freirich2021theory}
Dror Freirich, Tomer Michaeli, and Ron Meir.
\newblock A theory of the distortion-perception tradeoff in wasserstein space.
\newblock \emph{NeurIPS}, 2021.

\bibitem[Guo et~al.(2017)Guo, Pleiss, Sun, and Weinberger]{guo2017calibration}
Chuan Guo, Geoff Pleiss, Yu Sun, and Kilian~Q Weinberger.
\newblock On calibration of modern neural networks.
\newblock In \emph{ICML}, 2017.

\bibitem[He et~al.(2016)He, Zhang, Ren, and Sun]{he2016deep}
Kaiming He, Xiangyu Zhang, Shaoqing Ren, and Jian Sun.
\newblock Deep residual learning for image recognition.
\newblock In \emph{CVPR}, 2016.

\bibitem[Ho et~al.(2020)Ho, Jain, and Abbeel]{ho2020denoising}
Jonathan Ho, Ajay Jain, and Pieter Abbeel.
\newblock Denoising diffusion probabilistic models.
\newblock \emph{NeurIPS}, 2020.

\bibitem[Huang et~al.(2015)Huang, Singh, and Ahuja]{huang2015single}
Jia-Bin Huang, Abhishek Singh, and Narendra Ahuja.
\newblock Single image super-resolution from transformed self-exemplars.
\newblock In \emph{CVPR}, 2015.

\bibitem[Ji et~al.(2020)Ji, Cao, Tai, Wang, Li, and Huang]{ji2020real}
Xiaozhong Ji, Yun Cao, Ying Tai, Chengjie Wang, Jilin Li, and Feiyue Huang.
\newblock Real-world super-resolution via kernel estimation and noise injection.
\newblock In \emph{CVPRW}, 2020.

\bibitem[Ledig et~al.(2017)Ledig, Theis, Husz{\'a}r, Caballero, Cunningham, Acosta, Aitken, Tejani, Totz, Wang, et~al.]{ledig2017photo}
Christian Ledig, Lucas Theis, Ferenc Husz{\'a}r, Jose Caballero, Andrew Cunningham, Alejandro Acosta, Andrew Aitken, Alykhan Tejani, Johannes Totz, Zehan Wang, et~al.
\newblock Photo-realistic single image super-resolution using a generative adversarial network.
\newblock In \emph{CVPR}, 2017.

\bibitem[Li et~al.(2022)Li, Liu, Chen, Cai, Gu, Qiao, and Dong]{li2022blueprint}
Zheyuan Li, Yingqi Liu, Xiangyu Chen, Haoming Cai, Jinjin Gu, Yu Qiao, and Chao Dong.
\newblock Blueprint separable residual network for efficient image super-resolution.
\newblock In \emph{CVPR}, 2022.

\bibitem[Liang et~al.(2021)Liang, Cao, Sun, Zhang, Van~Gool, and Timofte]{liang2021swinir}
Jingyun Liang, Jiezhang Cao, Guolei Sun, Kai Zhang, Luc Van~Gool, and Radu Timofte.
\newblock Swinir: Image restoration using swin transformer.
\newblock In \emph{ICCV}, 2021.

\bibitem[Liang et~al.(2022)Liang, Zeng, and Zhang]{liang2022details}
Jie Liang, Hui Zeng, and Lei Zhang.
\newblock Details or artifacts: A locally discriminative learning approach to realistic image super-resolution.
\newblock In \emph{CVPR}, 2022.

\bibitem[Lim et~al.(2017)Lim, Son, Kim, Nah, and Mu~Lee]{lim2017enhanced}
Bee Lim, Sanghyun Son, Heewon Kim, Seungjun Nah, and Kyoung Mu~Lee.
\newblock Enhanced deep residual networks for single image super-resolution.
\newblock In \emph{CVPRW}, 2017.

\bibitem[Lu et~al.(2022)Lu, Li, Liu, Huang, Zhang, and Zeng]{lu2022transformer}
Zhisheng Lu, Juncheng Li, Hong Liu, Chaoyan Huang, Linlin Zhang, and Tieyong Zeng.
\newblock Transformer for single image super-resolution.
\newblock In \emph{CVPRW}, 2022.

\bibitem[Lugmayr et~al.(2020)Lugmayr, Danelljan, and Timofte]{lugmayr2020ntire}
Andreas Lugmayr, Martin Danelljan, and Radu Timofte.
\newblock Ntire 2020 challenge on real-world image super-resolution: Methods and results.
\newblock In \emph{CVPRW}, 2020.

\bibitem[Lugmayr et~al.(2021)Lugmayr, Danelljan, and Timofte]{lugmayr2021ntire}
Andreas Lugmayr, Martin Danelljan, and Radu Timofte.
\newblock Ntire 2021 learning the super-resolution space challenge.
\newblock In \emph{CVPR}, 2021.

\bibitem[Lyn(2020)]{lyn2020multi}
Jiawen Lyn.
\newblock Multi-level feature fusion mechanism for single image super-resolution.
\newblock \emph{arXiv preprint arXiv:2002.05962}, 2020.

\bibitem[Ma et~al.(2020)Ma, Rao, Cheng, Chen, Lu, and Zhou]{ma2020structure}
Cheng Ma, Yongming Rao, Yean Cheng, Ce Chen, Jiwen Lu, and Jie Zhou.
\newblock Structure-preserving super resolution with gradient guidance.
\newblock In \emph{CVPR}, 2020.

\bibitem[Ma et~al.(2019)Ma, Chu, Zhang, and Wan]{ma2019matrix}
Hailong Ma, Xiangxiang Chu, Bo Zhang, and Shaohua Wan.
\newblock A matrix-in-matrix neural network for image super resolution.
\newblock \emph{arXiv preprint arXiv:1903.07949}, 2019.

\bibitem[Mei et~al.(2021)Mei, Fan, and Zhou]{mei2021image}
Yiqun Mei, Yuchen Fan, and Yuqian Zhou.
\newblock Image super-resolution with non-local sparse attention.
\newblock In \emph{CVPR}, 2021.

\bibitem[Naeini et~al.(2015)Naeini, Cooper, and Hauskrecht]{naeini2015obtaining}
Mahdi~Pakdaman Naeini, Gregory Cooper, and Milos Hauskrecht.
\newblock Obtaining well calibrated probabilities using bayesian binning.
\newblock In \emph{AAAI}, 2015.

\bibitem[Neelakantan et~al.(2015)Neelakantan, Vilnis, Le, Sutskever, Kaiser, Kurach, and Martens]{neelakantan2015adding}
Arvind Neelakantan, Luke Vilnis, Quoc~V Le, Ilya Sutskever, Lukasz Kaiser, Karol Kurach, and James Martens.
\newblock Adding gradient noise improves learning for very deep networks.
\newblock \emph{arXiv preprint arXiv:1511.06807}, 2015.

\bibitem[Niu et~al.(2022)Niu, Zhu, Zhang, Sun, Wang, Kweon, and Zhang]{niu2022ms2net}
Axi Niu, Yu Zhu, Chaoning Zhang, Jinqiu Sun, Pei Wang, In~So Kweon, and Yanning Zhang.
\newblock Ms2net: Multi-scale and multi-stage feature fusion for blurred image super-resolution.
\newblock \emph{IEEE TCSVT}, 2022.

\bibitem[Niu et~al.(2023{\natexlab{a}})Niu, Wang, Zhu, Sun, Yan, and Zhang]{niu2023gran}
Axi Niu, Pei Wang, Yu Zhu, Jinqiu Sun, Qingsen Yan, and Yanning Zhang.
\newblock Gran: Ghost residual attention network for single image super resolution.
\newblock \emph{Multimedia Tools and Applications}, 2023{\natexlab{a}}.

\bibitem[Niu et~al.(2023{\natexlab{b}})Niu, Zhang, Pham, Wang, Sun, Kweon, and Zhang]{niu2023learning}
Axi Niu, Kang Zhang, Trung~X Pham, Pei Wang, Jinqiu Sun, In~So Kweon, and Yanning Zhang.
\newblock Learning from multi-perception features for real-word image super-resolution.
\newblock \emph{arXiv preprint arXiv:2305.18547}, 2023{\natexlab{b}}.

\bibitem[Park et~al.(2023)Park, Son, and Lee]{park2023content}
JoonKyu Park, Sanghyun Son, and Kyoung~Mu Lee.
\newblock Content-aware local gan for photo-realistic super-resolution.
\newblock In \emph{ICCV}, 2023.

\bibitem[Saharia et~al.(2022)Saharia, Ho, Chan, Salimans, Fleet, and Norouzi]{saharia2022image}
Chitwan Saharia, Jonathan Ho, William Chan, Tim Salimans, David~J Fleet, and Mohammad Norouzi.
\newblock Image super-resolution via iterative refinement.
\newblock \emph{TPAMI}, 2022.

\bibitem[Soh et~al.(2019)Soh, Park, Jo, and Cho]{soh2019natural}
Jae~Woong Soh, Gu~Yong Park, Junho Jo, and Nam~Ik Cho.
\newblock Natural and realistic single image super-resolution with explicit natural manifold discrimination.
\newblock In \emph{CVPR}, 2019.

\bibitem[Tashiro et~al.(2020)Tashiro, Song, and Ermon]{tashiro2020diversity}
Yusuke Tashiro, Yang Song, and Stefano Ermon.
\newblock Diversity can be transferred: Output diversification for white-and black-box attacks.
\newblock \emph{NeurIPS}, 2020.

\bibitem[Tian et~al.(2022)Tian, Zhang, Lin, Zuo, Zhang, and Lin]{tian2022generative}
Chunwei Tian, Xuanyu Zhang, Jerry Chun-Wei Lin, Wangmeng Zuo, Yanning Zhang, and Chia-Wen Lin.
\newblock Generative adversarial networks for image super-resolution: A survey.
\newblock \emph{arXiv preprint arXiv:2204.13620}, 2022.

\bibitem[Timofte et~al.(2017)Timofte, Agustsson, Van~Gool, Yang, and Zhang]{timofte2017ntire}
Radu Timofte, Eirikur Agustsson, Luc Van~Gool, Ming-Hsuan Yang, and Lei Zhang.
\newblock Ntire 2017 challenge on single image super-resolution: Methods and results.
\newblock In \emph{CVPRW}, 2017.

\bibitem[Wang et~al.(2018)Wang, Yu, Wu, Gu, Liu, Dong, Qiao, and Change~Loy]{wang2018esrgan}
Xintao Wang, Ke Yu, Shixiang Wu, Jinjin Gu, Yihao Liu, Chao Dong, Yu Qiao, and Chen Change~Loy.
\newblock Esrgan: Enhanced super-resolution generative adversarial networks.
\newblock In \emph{ECCVW}, 2018.

\bibitem[Wang et~al.(2021)Wang, Xie, Dong, and Shan]{wang2022realesrgan}
Xintao Wang, Liangbin Xie, Chao Dong, and Ying Shan.
\newblock Realesrgan: Training real-world blind super-resolution with pure synthetic data supplementary material.
\newblock \emph{ICCVW}, 2021.

\bibitem[Wang et~al.(2022)Wang, Wan, Yang, Li, Chau, and Kot]{wang2022low}
Yufei Wang, Renjie Wan, Wenhan Yang, Haoliang Li, Lap-Pui Chau, and Alex Kot.
\newblock Low-light image enhancement with normalizing flow.
\newblock In \emph{AAAI}, 2022.

\bibitem[Whang et~al.(2022)Whang, Delbracio, Talebi, Saharia, Dimakis, and Milanfar]{whang2022deblurring}
Jay Whang, Mauricio Delbracio, Hossein Talebi, Chitwan Saharia, Alexandros~G Dimakis, and Peyman Milanfar.
\newblock Deblurring via stochastic refinement.
\newblock In \emph{CVPR}, 2022.

\bibitem[Woo et~al.(2018)Woo, Park, Lee, and Kweon]{woo2018cbam}
Sanghyun Woo, Jongchan Park, Joon-Young Lee, and In~So Kweon.
\newblock Cbam: Convolutional block attention module.
\newblock In \emph{ECCV}, 2018.

\bibitem[Wu et~al.(2023)Wu, Jiang, and Liu]{wu2023practical}
Gang Wu, Junjun Jiang, and Xianming Liu.
\newblock A practical contrastive learning framework for single-image super-resolution.
\newblock \emph{TNNLS}, 2023.

\bibitem[Wu et~al.(2021)Wu, Qu, Lin, Zhou, Qiao, Zhang, Xie, and Ma]{wu2021contrastive}
Haiyan Wu, Yanyun Qu, Shaohui Lin, Jian Zhou, Ruizhi Qiao, Zhizhong Zhang, Yuan Xie, and Lizhuang Ma.
\newblock Contrastive learning for compact single image dehazing.
\newblock In \emph{CVPR}, 2021.

\bibitem[Xie et~al.(2023)Xie, Wang, Chen, Li, Shan, Zhou, and Dong]{xie2023desra}
Liangbin Xie, Xintao Wang, Xiangyu Chen, Gen Li, Ying Shan, Jiantao Zhou, and Chao Dong.
\newblock Desra: Detect and delete the artifacts of gan-based real-world super-resolution models.
\newblock 2023.

\bibitem[Xue et~al.(2023)Xue, Herranz, Corral, and Zhang]{xue2023burst}
Danna Xue, Luis Herranz, Javier~Vazquez Corral, and Yanning Zhang.
\newblock Burst perception-distortion tradeoff: Analysis and evaluation.
\newblock In \emph{ICASSP}, 2023.

\bibitem[Zeyde et~al.(2012)Zeyde, Elad, and Protter]{zeyde2012single}
Roman Zeyde, Michael Elad, and Matan Protter.
\newblock On single image scale-up using sparse-representations.
\newblock In \emph{International conference on curves and surfaces}, 2012.

\bibitem[Zhang et~al.(2018{\natexlab{a}})Zhang, Cisse, Dauphin, and Lopez-Paz]{zhang2018mixup}
Hongyi Zhang, Moustapha Cisse, Yann~N Dauphin, and David Lopez-Paz.
\newblock mixup: Beyond empirical risk minimization.
\newblock In \emph{ICLR}, 2018{\natexlab{a}}.

\bibitem[Zhang et~al.(2020)Zhang, Gool, and Timofte]{zhang2020deep}
Kai Zhang, Luc~Van Gool, and Radu Timofte.
\newblock Deep unfolding network for image super-resolution.
\newblock In \emph{CVPR}, 2020.

\bibitem[Zhang et~al.(2021)Zhang, Liang, Van~Gool, and Timofte]{zhang2021designing}
Kai Zhang, Jingyun Liang, Luc Van~Gool, and Radu Timofte.
\newblock Designing a practical degradation model for deep blind image super-resolution.
\newblock In \emph{ICCV}, 2021.

\bibitem[Zhang et~al.(2022)Zhang, Deng, Kawaguchi, and Zou]{zhang2022and}
Linjun Zhang, Zhun Deng, Kenji Kawaguchi, and James Zou.
\newblock When and how mixup improves calibration.
\newblock In \emph{ICML}, 2022.

\bibitem[Zhang et~al.(2018{\natexlab{b}})Zhang, Isola, Efros, Shechtman, and Wang]{zhang2018unreasonable}
Richard Zhang, Phillip Isola, Alexei~A Efros, Eli Shechtman, and Oliver Wang.
\newblock The unreasonable effectiveness of deep features as a perceptual metric.
\newblock In \emph{CVPR}, 2018{\natexlab{b}}.

\bibitem[Zhang et~al.(2018{\natexlab{c}})Zhang, Li, Li, Wang, Zhong, and Fu]{zhang2018image}
Yulun Zhang, Kunpeng Li, Kai Li, Lichen Wang, Bineng Zhong, and Yun Fu.
\newblock Image super-resolution using very deep residual channel attention networks.
\newblock In \emph{ECCV}, 2018{\natexlab{c}}.

\end{thebibliography}
}

\end{document}